\documentclass[11pt]{article}
\usepackage{amsmath} 

\usepackage{amscd}     \usepackage{amsxtra}
\usepackage{upref}     \usepackage{amsthm}
\usepackage{amssymb}   
\usepackage{amsfonts}
\begin{document}

\title{{\bf From Maxwell Stresses to Nonlinear Field Equations}}

\author{{\bf Stoil Donev}\footnote{e-mail:
 sdonev@inrne.bas.bg}, {\bf Maria Tashkova}, \\ Institute for Nuclear
Research and Nuclear Energy,\\ Bulg.Acad.Sci., 1784 Sofia,
blvd.Tzarigradsko chaussee 72\\ Bulgaria\\}

\date{}
\maketitle

\begin{abstract} This paper aims to show that making use of Newton's view
on equations of motion of a physical system and of the Maxwell stress
tensor we come to a natural nonlinearization of Maxwell equations in
vacuum making use only of nonrelativistic terms. The new equations include
all Maxwell solutions plus new ones, among which one may find time-stable
and spatially finite ones with photon-like properties and behavior.

\end{abstract}

\section{Introduction}
As it is well known the vacuum Maxwell equations (zero charge density:
$\rho=0$) do not admit spatially finite time-stable solutions of
photon-like type. This is due to the fact that every component
$U(x,y,z,t)$ of the electric {\bf E} and magnetic {\bf B} fields
necessarily satisfies the D'Alembert wave equation $\square U=0$, and
according to the Poisson's theorem for this equation, every spatially
finite initial condition $U(x,y,z,0)=\varphi(x,y,z);\  \frac{\partial
U}{\partial t}(x,y,z,0)=\psi(x,y,z)$, where $\varphi$ and $\psi$ are
finite functions, blows up radially and goes to infinity with the speed of
light [1,2,3]. So, through every spatial point outside the initial
condition pass fore-front and back-front, and after this the corresponding
point forgets about what has happened. This rigorous mathematical result
does not allow Maxwell vacuum equations to describe finite electromagnetic
pulses propagating uniformly in vacuum as a whole along some spatial
direction without dispersion. Moreover, no expectations for description of
photon-like objects having besides translational also rotational component
of propagation would seem to be reasonable and well-grounded.

On the other hand the Poynting energy-momentum balance equation in vacuum
$$
\frac{\partial}{\partial t}\left(\frac{\mathbf{E}^2+\mathbf{B}^2}{2}\right)=
-c\,\mathrm{div}(\mathbf{E}\times\mathbf{B}),
$$
where $c$ is the velocity of light in vacuum, admits time-stable spatially
finite solutions with 3-dimensional soliton-like behavior, for
example (no rotation component of propagation)
$$
\mathbf{E}=[u(x,y,ct+\varepsilon z),\,p\,(x,y,ct+\varepsilon z),0],\ \ \
\mathbf{B}=[\varepsilon p\,(x,y,ct+\varepsilon z),\,
-\varepsilon u(x,y,ct+\varepsilon z),0],  \ \ \varepsilon=\pm1\ ,
$$
where $u$ and $p$ are {\it arbitrary} functions, so they can be chosen
{\it finite}. This observation suggests to look deeper and more carefully
into the structures and assumptions used for mathematical interpretation
of the experimental electric-magnetic induction discoveries made in the
19th century. In other words, which relations and on what grounds should
be defined as basic, so that the further deduced equations and relations
to give reasonable and physically acceptable results. Finding the right
way to choosing adequate mathematical objects and corresponding equations
seems specially important when we try to describe the intrinsic dynamical
properties of such spatially finite time stable field objects. Therefore,
it seems very important to have the right notion about concepts like
physical object, intrinsic property, dynamical property, identifying
characteristics, admissible changes, field equations, etc. Some
preliminary considerations in this direction might be the following ones.

From a definite point of view every physical system is characterized by
two kinds of properties.  The {\bf first} kind of properties we call {\it
identifying}, they identify the system throughout its existence in time,
so, the corresponding physical quantities/relations must show definite
conservation/constancy properties (with respect to the identification
procedure assumed). Without such experimentally established properties we
could not talk about physical objects/systems at all.  The {\bf second}
kind of properties (which may be called {\it kinematical}) characterize the
time-evolution of the system, the corresponding quantities are time-dependent,
and the corresponding evolution is {\it consistent} with the
conservative/constant character of the identifying properties/quantities. In
this sense, the equations of motion of a physical system can be considered as
relations determining the admissible time-changes of these quantities. For
example, the mass $m$ of a classical particle is an identifying quantity, while
the velocity $\mathbf v$ is a kinematical one. This view implies, of course,
that the external world acts on the system under consideration also in an
admissible way, i.e. an assumption is made that the system survives, the
interaction with the outside world does not lead to its destruction.

In theoretical physics we usually make use of quantities which are
functions of the identifying and of the kinematical characteristics of the
system and call them {\it dynamical} quantities. A well known example is
the momentum $\mathbf{p}$ of a particle: $\mathbf p=m\mathbf v$. Of
crucial importance for the description of admissible changes are the
conservative dynamical quantities, i.e.  those which may pass from one
physical system to another with NO LOSS.  For example {\bf energy} and
{\bf momentum} are such quantities, moreover, they are universal in the
sense that every physical object carries non-zero energy-momentum and,
vice versa, every quantity of energy-momentum is carried by some physical
object. So, if a definite quantity of energy-momentum passes from one
object to another, this same quantity of energy-momentum can be expressed
in terms of the characteristics of the two objects/systems, and the two
expressions to be equalized. In this way we obtain consistent with the
energy-momentum conservation law equations of motion, and this is the way
used by Newton to write down his famous equations $\dot{\mathbf p}
=\mathbf F $, where $\mathbf F$ carries information about where the
momentum change of the particle has gone, or has come from.  This also
clarifies {\it the physical sense of the concept of force as a change of
momentum}, or as a change of energy-momentum in relativistic terms. Paying
due respect to Newton we shall call some equations of motion of {\it
Newton type} if on the two sides of $"="$ stay physical quantities of
energy-momentum change, or energy-momentum density change in the case of
continuous systems. Note that, written down for the vector field
$\mathbf{p}$, i.e. in terms of partial derivatives, the above Newton
equation looks like $\nabla_{\mathbf{p}}\mathbf{p}=m\mathbf{F}$, where the
left hand side means performing two steps: first, determining the "change
quantity" $\nabla{\mathbf{p}}$, second, projecting $\nabla{\mathbf{p}}$ on
$\mathbf{p}$, and the right hand side may be expressed as a function of
the characteristics of both: the particle and the external physical
envirenment.

If there is no energy-momentum (or energy-momentum density) change, then
putting the corresponding expression equal to zero, e.g.
$\nabla_{\mathbf{p}}{\mathbf{p}}=0$, we obtain the "free particle" or
"free field" equations. In such a case we just declare that only those
changes are admissible which are consistent with the (local and integral)
energy-momentum conservation.

We note that an initial extent of knowledge about the system we are going
to describe mathematically is presupposed to be available, so that the
assumptions made to be, more or less, well grounded. This knowledge is the
base that generates corresponding insight and directs our attention to the
appropriate mathematical structures. This is exclusively important when we
deal with continuous, or field, physical objects/systems.

In view of the above considerations, roughly speaking, in the {\it free
field} case the steps to follow are:

1. Specify and consider the mathematical model-object $\Phi$ which is
chosen to represent the integrity of the physical system considered;

2. Define the change-object $D(\Phi)$;

3. "Project" $D(\Phi)$ on $\Phi$ by means of some (in most cases bilinear)
map $\mathfrak{P}$;

4. The projection $\mathfrak{P}(D(\Phi),\Phi)$ obtained we interpret
physically as energy-momentum change;

5. We put this projection equal to zero: $\mathfrak{P}(D(\Phi),\Phi)=0$ .
\vskip 0.3cm The zero value of the projection $\mathfrak{P}(D(\Phi),\Phi)$
is interpreted in the sense that the identifying characteristics of $\Phi$
have not been disturbed, or, the change $D(\Phi)$ is qualified as {\it
admissible}. This consideration shows the importance of knowing how much
and in what way(s) a given physical system is {\it potentially able} to
lose, or gain energy-momentum (locally or globally), without losing its
identity.

It is always very important to take care of the physical sense of the
quantities that we put on the two sides of the relation $A=B$.
Mathematically, from set theory point of view, $A$ and $B$ denote {\it the
same} element, which element may be expressed in different terms, e.g. the
real number 2 can be expressed as
$3-1=6/3=2(\mathrm{sin}^2x+\mathrm{cos}^2x)=\frac{d}{dx}(2x+const)$ and
also in many other ways. From physical point of view, however, we must be
sure that $A$ and $B$ denote the same thing {\it qualitatively} and {\it
quantitatively}, i.e. the same {\it physical} quantity. This is specially
important when the equation we want to write down constitutes some basic
relation. And the point is not the physical dimension of the two sides to
be the same: any two quantities by means of an appropriate constant can be
made of the same physical dimension, but this is a formal step. The point
is that {\it the physical nature of the physical quantity on the two sides
must be the same}.

For example, it is quite clear that on the two sides of the Newton's law
$\dot{\mathbf p} =\mathbf F $ stays the well defined for any physical system
quantity "change of momentum" since the {\it momentum} quantity is a
universal one. For a counterexample, which physical quantity stays on the
two sides of the Poisson equation $\Delta U=k\rho$?  On one hand, such a
quantity is expressed through $\Delta U$ and, since $grad\,U$ is usually
interpreted as force, $\Delta U$ appears as a "change of force"
characteristic of the field $U$ since it is essentially defined by the
second derivatives of $U$. On the other hand, the same quantity is expressed
through $k\rho$ and appears as a characteristic of the mass particles, so,
do we know such a quantity? The same question can be raised for one of the
Maxwell equations:
$\mathrm{rot}\,\mathbf{B}-\frac1c\dot{\mathbf{E}}=\frac{4\pi}{c}\mathbf{j}$.

In the case of classical particles momentum is always represented as the
product $m\mathbf v$ and this is carried to fluid mechanics as
$\mu(x,y,z;t).\mathbf v(x,y,z;t)$, where $\mu $ is the mass density. A
similar quantity is introduced in electrodynamics as electric current
density $\mathbf{j}=\rho(x,y,z;t).\mathbf v(x,y,z;t)$, where $\rho$ is the
electric charge density. The energy-momentum exchange between the field and
the available charged particles is described by the force field
$\mathbf{F}=\rho\mathbf{E} + \frac{1}{c}\mathbf{j}\times\mathbf{B}$. So, the
corresponding Faraday-Maxwell {\it force lines} should be the integral lines
of the vector field $\mathbf{F}$. Clearly, in the {\it charge-free} case we
get $\mathbf{F}=0$, so {the concept of force-lines defined by $\mathbf{F}$
does not work}. Hence, if we would like to use this concept appropriately in
the charge-free case, we have to introduce it appropriately. The simplest
way seems to consider the integral lines of $\mathbf{E}$ and $\mathbf{B}$ as
force lines also in the charge free case, but we do not share this view: if
$\rho=0$ then $\mathbf{j}=0$, the force-vector is zero and NO integral force
lines exist. The two vectors $\mathbf{E}$ and $\mathbf{B}$ generate, of
course, integral lines, but these integral lines are NOT force lines since
$\mathbf{E}$ and $\mathbf{B}$ are NOT force fields, and such an
interpretation of the integral lines of $\mathbf{E}$ and $\mathbf{B}$ would
be misleading. In fact, there exist a {\it sufficiently good} force field
defined by Maxwell in terms of the divergence of his stress tensor $M^{ij}$,
which definition {\bf works quite well also out of and away from any media
built of, or containing, charged mass particles}. So, in the frame of the
theory at the end of 19th century if we ask the question: {\it if there are
NO charged particles and the time-dependent EM-field cannot transfer
energy-momentum to them by means of the force field $\mathbf{F}$, and the
propagation of the free EM-field is available, so that energy-momentum
internal exchanges should necessarily take place, how these processes and
the entire propagational behaviour of the field could be understood and
modeled}?, the right answer in our view should be: {\bf turn to $M^{ij}$ and
consider carefully the divergence $\nabla_iM^{ij}$ as corresponding force
field generating corresponding force lines along which energy-momentum is
locally transported}. As will be seen further in the paper, such a look on
the issue would necessarily lead Maxwell and his followers to the prediction
that real, free, spatially finite and time-stable formations of
electromagnetic field nature having translational-rotational dynamical
structure should exist, a result that has been proved in studying the
photoeffect phenomena about 30 years after Maxwell's death.

We consider as a remarkable achievement of Maxwell the determination of
the correct expressions for the energy density of the electromagnetic field
through the concept of {\it stress} [4]. His electromagnetic {\it stress
tensor} $M^{ij}$ still plays an essential role in modern electromagnetic
theory as a part of the modern relativistic stress-energy-momentum tensor.
However, by some reasons, Maxwell did not make further use of the computed
by him divergence $\nabla_iM^{ij}$ of the stress tensor (and called by him
"force field" [4]) for writing down Newton type equations of motion for a
free electromagnetic field through {\it equalizing different expressions
for the same momentum change}. Probably, he had missed an appropriate
interpretation of the vector $c\,\mathbf{E}\times\mathbf{B}$ (introduced by
Poynting 5 years after Maxwell's death and called "electromagnetic energy
flux" [5]).

In this paper we consider one possible approach to come to natural {\it free
field} equations of motion of Newton type that could be deduced making use
of formally introduced vacuum analog of the Maxwell stress tensor and of the
Poynting vector {\it in the frame of the available theoretical notions and
concepts at the end of 19th century}. As a first step we are going to show
that an analog of Maxwell's stress tensor participates in a (well known
today) mathematical identity having nothing to do with any physics.

\section{A non-physical view on Maxwell stress tensor} The mathematical
identities have always attracted the attention of theorists, in
particular, those identities which involve the derivatives of the objects
of interest (differential identities). A well known such example is the
Bianchi identity satisfied by any connection components: this identity is
a second order system of (in general, nonlinear) partial differential
equations. The gauge interpretation of classical Maxwell electrodynamics,
as well as the Yang-Mills theory, substantially make use of this identity.
Such identities are of particular importance when on the two sides of "="
stay correctly (i.e. in a coordinate free way) defined expressions.

It is elementary to show that in the frame of classical vector analysis
any two vector fields $(V,W)$ and the corresponding mathematical analog of
the Maxwell stress tensor $M^{ij}(V,W)$ are involved in a differential
identity. Introducing the Maxwell stress tensor in such a formal way in
the {\it vacuum case} will help us avoid all questions concerning the
structure and properties of {\it aether}.

We begin with the well known differential relation satisfied by every
vector field $V$ on the euclidean space $\mathbb{R}^3$ related to the
standard coordinates $(x^i=x,y,z), i=1,2,3$, denoting by $V^2$ the
euclidean square of $V$, by $"\times"$ - the vector product, and using the
$\nabla$-operator: \[ \frac12\nabla(V^2)=V\times
\mathrm{rot}\,V+(V.\nabla)V =V\times \mathrm{rot}\,V + \nabla_V V. \]
Clearly, on the two sides of this relation stay well defined quantities,
i.e. quantities defined in a coordinate free way. The first term on the
right hand side of this identity accounts for the rotational component of
the change of $V$, and the second term accounts mainly for the
translational component of the change of $V$. Making use of component
notation we write down the last term on the right side as follows
(summation over the repeated indices):
\[
(\nabla_V V)^j=V^i\nabla_i
V^j=\nabla_i(V^iV^j)-V^j\nabla_iV^i= \nabla_i(V^iV^j)-V^j\mathrm{div}\,V .
\]
Substituting into the first identity, and making some elementary
transformations we obtain
\[
\nabla_i\left(V^iV^j-\frac12
\delta^{ij}V^2\right)= \big [(\mathrm{rot}\,V)\times
V+V\mathrm{div}\,V\big ]^j,
\]
where $\delta^{ij}=1$ for $i=j$, and
$\delta^{ij}=0$ for $i\neq j$. If now $W$ is another vector field it must
satisfy the same above identity:
\[
\nabla_i\left(W^iW^j-\frac12
\delta^{ij}W^2\right)= \big [(\mathrm{rot}\,W)\times
W+W\mathrm{div}\,W\big ]^j.
\]
Summing up these two identities we obtain
the new identity
\setlength\arraycolsep{8pt}
\begin{eqnarray}
\lefteqn{
\nabla_iM^{ij}\equiv \nabla_i\left(V^iV^j+W^iW^j-
\delta^{ij}\frac{V^2+W^2}{2}\right)={} } \nonumber\\ & & {}=\big
[(\mathrm{rot}\,V)\times V+ V\mathrm{div}\,V+(\mathrm{rot}\,W)\times
W+W\mathrm{div}\,W\big ]^j.
\end{eqnarray}
We emphasize once again the two moments: first, this identity (1) has
nothing to do with any physics; second, on the two sides of (1) stay well
defined coordinate free quantities. We note also the invariance of
$M^{ij}$ with respect to the transformations $(V,W)\rightarrow (-W,V)$ and
$(V,W)\rightarrow (W,-V)$.

The expression inside the round brackets on the left of (1), denoted by
$M^{ij}$, looks formally the same as the introduced by Maxwell tensor from
physical considerations concerned with the electromagnetic stress energy
properties of continuous media in presence of external electromagnetic
field. This allows to call formally any such tensor {\bf Maxwell stress
tensor} generated by the two vector fields $(V,W)$ . The term "stress" in
this general mathematical setting could be interpreted (or, justified) in
the following way. Every vector field on $\mathbb{R}^3$ generates
corresponding flow by means of the trajectories started from some domain
$U_o\subset\mathbb{R}^3$:  at the moment $t>0$ the domain $U_o$ is
diffeomorphically transformed to a new domain $U_t\subset\mathbb{R}^3$.
Having two vector fields on $\mathbb{R}^3$ we obtain two {\it consistent}
flows, so, the points of any domain $U_o\subset\mathbb{R}^3$ are forced to
accordingly move to new positions.

Physically, we say that the corresponding physical medium that occupies
the spatial region $U_o$ and is parametrized by the points of the
mathematical subregion $U_o\subset\mathbb{R}^3$, is subject to {\it
consistent} and {\it admissible} physical "stresses" generated by physical
interactions mathematically described by the couple of vector fields
$(V,W)$, and these physical stresses are quantitatively described by the
corresponding physical interpretation of the tensor $M^{ij}(V,W)$.

We note that the stress tensor $M^{ij}$ in (1) is subject to the divergence
operator, and if we interpret the components of $M^{ij}$ as physical
stresses, then the left hand side of (1) acquires in general the physical
interpretation of force density. Of course, in the static situation as it is
given by relation (1), no energy-momentum propagation is possible, so at
every point the forces mutually compensate: $\nabla_{i}M^{ij}=0$. If
propagation is allowed then the force field is NOT zero:
$\nabla_{i}M^{ij}\neq 0$, and we may identify the right hand side of (1) as
a {\bf real time-change} of appropriately defined momentum density
$\mathbf{P}$. So, assuming some expression for this momentum density
$\mathbf{P}$ we are ready to write down corresponding field equation of
motion of Newton type through equalizing the spatially directed force
densities $\nabla_{i}M^{ij}$ with the momentum density changes along the
time coordinate, i.e. equalizing $\nabla_iM^{ij}$ with the $ct$-derivative
of $\mathbf{P}$, where $c=const$ is the translational propagation velocity
of the momentum density flow of the physical system $(V,W)$. In order to
find how to choose $\mathbf{P}$ we consider briefly the eigen properties of
$M^{ij}$.

\section{Eigen properties of the Maxwell stress tensor}
We consider $M^{ij}$ at some point of $\mathbb{R}^3$ and assume that in
general the vector fields $\mathbf{E}$ and $\mathbf{B}$ are linearly
inependent, so $\mathbf{E}\times\mathbf{B}\neq 0$. Let the coordinate system
be chosen such that the coordinate plane $(x,y)$ to coinside with the plane
defined by $\mathbf{E},\mathbf{B}$. In this coordinate system
$\mathbf{E}=(E_1,E_2,0)$ and $\mathbf{B}=(B_1,B_2,0)$, so, identifying the
contravariant and covariant indices through the Euclidean metric (so that
$M^{ij}=M^i_j=M_{ij}$), we obtain the following nonzero components of the
stress tensor: $$ M^1_1=(E^1)^2+(B^1)^2-\frac12(\mathbf{E}^2+\mathbf{B}^2);
\ \ M^1_2=M^2_1=E^1\,E_2+B_1\,B^2; $$ $$
M^2_2=(E^2)^2+(B^2)^2-\frac12(\mathbf{E}^2+\mathbf{B}^2); \ \
M^3_3=-\frac12(\mathbf{E}^2+\mathbf{B}^2). $$ Since $M^1_1=-M^2_2$, the
trace of $M$ is $Tr(M)=-\frac12(\mathbf{E}^2+\mathbf{B}^2)$. The eigen value
equation acquires the simple form
$\big[(M^1_1)^2-(\lambda)^2\big]+(M^1_2)^2\big](M^3_3-\lambda)=0$. The
corresponding eigen values are $$
\lambda_1=-\frac12(\mathbf{E}^2+\mathbf{B}^2);\ \
\lambda_{2,3}=\pm\sqrt{(M^1_1)^2+(M^1_2)^2}=
\pm\frac12\sqrt{(I_1)^2+(I_2)^2} , $$ where
$I_1=\mathbf{B}^2-\mathbf{E}^2,\, I_2=2\mathbf{E}.\mathbf{B}$. The
corresponding to $\lambda_1$ eigen vector $Z_1$ must satisfy the equation
$\mathbf{E}(\mathbf{E}.Z_1)+\mathbf{B}(\mathbf{B}.Z_1)=0$, hence, $Z_1$ is
proportional to $\mathbf{E}\times\mathbf{B}$:
$Z_1=k\,\mathbf{E}\times\mathbf{B}$. The other two eigen vectors $Z_{1,2}$
satisfy correspondingly the equations $$ \mathbf{E}(\mathbf{E}.Z_{1,2})+
\mathbf{B}(\mathbf{B}.Z_{1,2})=\Big[\pm\frac12\sqrt{(I_1)^2+(I_2)^2}+
\frac12(\mathbf{E}^2+\mathbf{B}^2)\Big]Z_{1,2} . $$ Taking into account the
relation $$ \frac14\Big[(I_1)^2+(I_2)^2\Big]=
\left(\frac{\mathbf{E}^2+\mathbf{B}^2}{2}\right)^2-
|\mathbf{E}\times\mathbf{B}|^2 \ , \ \ \text{so}\ \ \ \,
\frac{\mathbf{E}^2+\mathbf{B}^2}{2}- |\mathbf{E}\times\mathbf{B}|\geq 0 \ ,
$$ we conclude that the coefficient before $Z_{1,2}$ on the right is always
different from zero, therefore, the eigen vectors $Z_{1,2}$ lie in the plane
defined by $(\mathbf{E},\mathbf{B})$.

The above consideration suggests that {\it the intrinsically defined potential
dynamical abilities of propagation of the field are: translational along
$(\mathbf{E}\times\mathbf{B})$, and rotational inside the plane defined by}
$(\mathbf{E},\mathbf{B})$.

We give some motivation now that the theory should choose
$\lambda_2=\lambda_3=0$. In fact, it turns out that if these two relations
do not hold then the translational velocity of propagation is allowed to be
less then the speed of light in vacuum $c$. Recall first the transformation
laws of the electric and magnetic vectors under Lorentz transformation
defined by the 3-velocity vector $\mathbf{v}$ and corresponding parameter
$\beta=v/c$. If $\gamma$ denotes the factor $1/\sqrt{1-\beta^2}$ then we
have $$
\mathbf{E'}=\gamma\,\mathbf{E}+\frac{1-\gamma}{v^2}\mathbf{v}(\mathbf{E}.
\mathbf{v})+\frac{\gamma}{c}\mathbf{v}\times\mathbf{B} , $$ $$
\mathbf{B'}=\gamma\,\mathbf{B}+\frac{1-\gamma}{v^2}\mathbf{v}(\mathbf{B}.
\mathbf{v})-\frac{\gamma}{c}\mathbf{v}\times\mathbf{E} . $$

Assume first that $I_2=2\mathbf{E}.\mathbf{B}=0$, i.e.
$\mathbf{E}$ and $\mathbf{B}$ are orthogonal, so, in general, in some
coordinate system we shall have  $\mathbf{E}\times\mathbf{B}\neq 0$ .

If $I_1>0$, i.e. $|\mathbf{E}|<|\mathbf{B}|$. We shall show that the choice
$\frac{v}{c}=(|\mathbf{E}|/|\mathbf{B}|)<1$ is addmissible. Actually, the
assumptions $\mathbf{E'}=0$ and $\mathbf{v}.\mathbf{B}=0$ lead to
$\gamma\,\mathbf{v}.\mathbf{E}+(1-\gamma)(\mathbf{E}.\mathbf{v})=0$, i.e.
$\mathbf{E}.\mathbf{v}=0$. Thus, $\mathbf{E'}=0$ leads to $c|\mathbf{E}|=
v|\mathbf{B}||\mathrm{sin}(\mathbf{v},\mathbf{E})|$, and since
$\mathbf{v}.\mathbf{E}=0$ then $|\mathrm{sin}(\mathbf{v},\mathbf{E})|=1$. It
follows that the speed $v=c\frac{|\mathbf{E}|}{|\mathbf{B}|}<c$ is allowed,
so the vector product $\mathbf{E'}\times\mathbf{B'}=0$.

If $I_1<0$, i.e. $|\mathbf{E}|>|\mathbf{B}|$, then the choice
$\mathbf{B'}=0$ and $\mathbf{v}.\mathbf{E}=0$ analogically leads to the
conclusion that the speed $v=c\frac{|\mathbf{B}|}{|\mathbf{E}|}<c$ is
allowed and $\mathbf{E'}\times\mathbf{B'}=0$.

Assume now that $I_2=2\mathbf{E}.\mathbf{B}\neq 0$. We are looking for a
reference frame $K'$ such that $\mathbf{E'}\times\mathbf{B'}=0$, while in
the reference frame $K$ we have $\mathbf{E}\times\mathbf{B}\neq 0$. We
choose the relative velocity $\mathbf{v}$ such that $\mathbf{v}.\mathbf{E}=
\mathbf{v}.\mathbf{B}=0$. Under these conditions the equation
$\mathbf{E'}\times\mathbf{B'}=0$ reduces to $$
\mathbf{E}\times\mathbf{B}+\frac{\mathbf{v}}{c}(\mathbf{E}^2+\mathbf{B}^2)=0
,\ \ \text{so}, \ \
\frac{v}{c}=|\mathbf{E}\times\mathbf{B}|/(\mathbf{E}^2+\mathbf{B}^2). $$
Now, from the above mentioned inequality
$\mathbf{E}^2+\mathbf{B}^2-2|\mathbf{E}\times\mathbf{B}|\geq 0$ it follows
that $\frac{v}{c}<1$.

These considerations show that under nonzero $I_1$ and $I_2$ the
translational velocity of propagation of the field, and of the energy
density of course, will NOT be equal to $c$. Hence, the only realistic
choice left is $I_1=I_2=0$, which is equivalent to
$\mathbf{E}^2+\mathbf{B}^2=2|\mathbf{E}\times\mathbf{B}|$. Hence, assuming
$|Tr(M)|$ to be the energy density of the field, the name "electromagnetic
energy flux" for the quantity $c\mathbf{E}\times\mathbf{B}$ seems well
justified without turning to any field equations.

These considerations show also that if $I_1=|\mathbf{B}|^2-|\mathbf{E}|^2=0$
then the electric and magnetic components of the field carry always the same
energy density, so, a local mutual energy exchange is not forbidden in
general, but, if it takes place, it must be {\it simultanious} and in {\it
equal quantities}. Hence, under zero invariants $I_1=0$ and
$I_2=2\mathbf{E}.\mathbf{B}=0$ internal energy exchange is allowed but this
exchange occurs {\it without available interaction energy}, and the field
equations must take care of such a possibility.

\section{Nonlinear equations for the electromagnetic field}
We procede to write down dynamical equations of the field through
specializing how the internal local momentum exchange is realized.

We note first that the assumption that the energy density of the field
coincides with $|Tr(M)|=\frac12[\mathbf{E}^2+\mathbf{B}^2]$ presupposes that
{\it there is NO interaction energy} between the electric and magnetic
components of the field. It is an usual practice to introduce interaction
energy through some function of the scalar product of the two fields:
$E_{int}=f(\mathbf{E}.\mathbf{B})$, e.g., proportional to
$\mathbf{E}.\mathbf{B}$. Hence, in the free field case it seems natural to
assume $\mathbf{E}.\mathbf{B}=0$, i.e. $I_2=0$. We just note that, this does
not mean that there is no energy exchange between the electric and magnetic
components.

Now, what about a possible {\it momentum} exchange between the electric and
magnetic components of the field? Note that the assumption that the field
momentum is given by $\frac1c\mathbf{E}\times\mathbf{B}$, i.e. it is a
bilinear function of the electric and magnetic components, clearly suggests
that there is available nonzero {\it interaction momentum}. The point now is
to get some clarification how this local momentum exchange takes place and
to find appropriate mathematical representatives of the corresponding
partners realizing such an exchange since neither $\mathbf{E}$ nor
$\mathbf{B}$ are able to carry momentum separately.

We procede to answer this question through writing down corresponding field
equations.

Replacing $(V,W)$ in (1) with $(\mathbf{E},\mathbf{B})$ we obtain
\setlength\arraycolsep{8pt}
\begin{eqnarray}
\lefteqn{ \nabla_iM^{ij}\equiv
\nabla_i\left(\mathbf{E}^i\mathbf{E}^j+\mathbf{B}^i\mathbf{B}^j-
\delta^{ij}\frac{\mathbf{E}^2+\mathbf{B}^2}{2}\right)={} }
\nonumber\\
& & {}=
\big [(\mathrm{rot}\,\mathbf{E})\times \mathbf{E}+
\mathbf{E}\mathrm{div}\,\mathbf{E}+
(\mathrm{rot}\,\mathbf{B})\times \mathbf{B}+
\mathbf{B}\mathrm{div}\,\mathbf{B}\big ]^j.
\end{eqnarray}

As we mentioned, in the static case, i.e. when the vector fields
$(\mathbf{E},\mathbf{B})$ do not depend on the time coordinate $\xi=ct$,
NO propagation of field momentum density $\mathbf{P}$ should take place,
so, at every point, where $(\mathbf{E},\mathbf{B})\neq 0$, the  stress
generated forces must mutually compensate, i.e. the divergence
$\nabla_iM^{ij}$ should be equal to zero: $\nabla_iM^{ij}=0$. In this
static case Maxwell vacuum equations
\[
\mathrm{rot}\,\mathbf{E}+\frac{\partial\mathbf{B}}{\partial \xi}=0,\quad
\mathrm{rot}\,\mathbf{B}-\frac{\partial\mathbf{E}}{\partial \xi}=0,\quad
\mathrm{div}\,\mathbf{E}=0,\quad \mathrm{div}\,\mathbf{B}=0 \ \ \ \ \ \ \
(*)
\]
give: $\mathrm{rot}\mathbf{E}=\mathrm{rot}\mathbf{B}=0;\,
\mathrm{div}\mathbf{E}=\mathrm{div}\mathbf{B}=0$, so, all static solutions
to Maxwell equations determine a sufficient, but NOT necessary, condition
that brings to zero the right hand side of (2) through forcing each of the
four vectors there to get zero values.

In the non-static case, i.e. when $\frac{\partial\mathbf{E}}{\partial t}\neq
0; \,\frac{\partial\mathbf{B}}{\partial t}\neq 0$, time change and
propagation of field momentum density should take place, so, a full mutual
compensation of the generated by the Maxwell stresses at every spatial point
local forces may NOT be possible, which means $\nabla_iM^{ij}\neq 0$ in
general. These local forces generate time-dependent momentum propagation
$\mathbf{P}(\mathbf{E},\mathbf{B})$ at the spatial points. Therefore, if we
want to describe this physical process of field momentum density time
change and spatial propagation we have to introduce explicitly the
dependence $\mathbf{P}(\mathbf{E},\mathbf{B})$. If we follow the classical
(nonrelativistic) way of consideration and denote by $\mathfrak{F}$ the
vector field with components $\mathfrak{F}^j=\nabla_iM^{ij}$, we can write
down the {\it force flow} across some finite 2-surface $S$ in the usual (and
widely spread in almost all textbooks) way as
$\int_{S}\mathfrak{F}.\mathbf{ds}$ (from modern point of view we should
write $i_{\mathfrak{F}}(dx\wedge dy\wedge dz)$ instead of
$\mathfrak{F}.\mathbf{ds}$ under the integral, where $i_{\mathfrak{F}}$
denotes the inner product between the vector field $\mathfrak{F}$ and the
volume form $dx\wedge dy\wedge dz$, i.e. to make use of the Poincare
isomorphism between vector fields and 2-forms on $\mathbb{R}^3$). This flow
generates changes of the momentum density flow across $S$ which should be
equal to
$\frac{d}{dt}\int_{S}\mathbf{P}(\mathbf{E},\mathbf{B}).\mathbf{ds}$. We
obtain \[ \frac{d}{dt}\int_{S}\mathbf{P}(\mathbf{E},\mathbf{B}).\mathbf{ds}=
\int_{S}\mathfrak{F}.\mathbf{ds} \ . \]
\vskip 0.3cm
The explicit expression
for  $\mathbf{P}(\mathbf{E},\mathbf{B})$, paying due respect to J.Poynting
[5], and to J.J.Thomson, H.Poincare, M. Abraham [6], and in view of the
more than a century development of the electricity theory and practice,
has to be introduced by the following
\vskip 0.4cm \noindent {\bf Assumption}: {\it The field momentum density is
given by $\mathbf{P}:=\frac1c\mathbf{E}\times\mathbf{B}$} .
 \vskip 0.4cm
According to the {\bf Assumption} and the above interpretation of the
relation $\nabla_iM^{ij}\neq 0$, and in view of the arbitrariness of the
2-surface $S$ we come to the vector differential equation
\[
\frac{\partial}{\partial
\xi}\left(\mathbf{E}\times\mathbf{B}\right)=\mathfrak{F},
\ \ \ \xi\equiv ct,   \ \ \ \ \ \ \ (**)
\]
which according to relation (2) is equivalent to
\begin{equation}
\left(\mathrm{rot}\,\mathbf{E}+\frac{\partial\mathbf{B}}{\partial
\xi}\right)\times \mathbf{E}+ \mathbf{E}\mathrm{div}\,\mathbf{E}+    
\left(\mathrm{rot}\,\mathbf{B}-\frac{\partial\mathbf{E}}{\partial
\xi}\right)\times \mathbf{B}+ \mathbf{B}\mathrm{div}\,\mathbf{B}=0.
\end{equation}
This last equation (3) we write down in the following equivalent way:
\begin{equation}
\left(\mathrm{rot}\,\mathbf{E}+\frac{\partial\mathbf{B}}{\partial
\xi}\right)\times \mathbf{E}+\mathbf{B}\mathrm{div}\,\mathbf{B}=     
-\left[\left(\mathrm{rot}\,\mathbf{B}-\frac{\partial\mathbf{E}}{\partial
\xi}\right)\times\mathbf{B}+\mathbf{E}\mathrm{div}\,\mathbf{E}\right].
\end{equation}
The above relation (**) and the corresponding differential relation
(3)/(4) we consider as mathematical adequate in momentum-change terms of
the electric-magnetic and magnetic-electric induction phenomena in the
charge free case. We recall that these induction phenomena are described
in what we call "Faraday-Maxwell theory" by the following well known
integral and differential equations
\[
\frac{d}{d\xi}\int_{S}\mathbf{B}.\mathbf{ds}=-
\int_{S}\mathrm{rot}\mathbf{E}.\mathbf{ds}\ \ \ \rightarrow \ \ \
\frac{\partial\mathbf{B}}{\partial \xi}=-\mathrm{rot}\mathbf{E}, \ \ \
\text{(the Faraday induction law)},
\]
\[
\frac{d}{d\xi}\int_{S}\mathbf{E}.\mathbf{ds}=
\int_{S}\mathrm{rot}\mathbf{B}.\mathbf{ds}\ \ \ \rightarrow \ \ \
\frac{\partial\mathbf{E}}{\partial \xi}=\mathrm{rot}\mathbf{B},  \ \ \
\text{(the Maxwell displacement current law)}.
\]
We stress once again that these last Faraday-Maxwell relations have NO {\it
direct} energy-momentum change-propagation (i.e. force flow) nature, so
they could not be experimentally verified in a {\it direct} way.
Our feeling is that, in fact, they are stronger than needed. So, on the
corresponding solutions of these equations we'll be able to write down
{\it formally adequate} energy-momentum change expressions, but the
consistency of these expressions with the experiment will crucially depend
on the nature of these solutions. As we already mentioned, the nature of
the free solutions (with no boundary conditions) to Maxwell vacuum
equations with spatially finite initial conditions requires strong
time-instability (the Poisson theorem for the D'Alembert wave equation).
And time-stability of time-dependent vacuum solutions usually requires spatial
infinity (plane waves), which is physically senseless. Making calculations
with spatially finite parts of these spatially infinite solutions may be
practically acceptable, but from theoretical viewpoint assuming
these equations for {\it basic} ones seems not acceptable since the relation
"time stable physical object - exact free solution" is strongly violated.

Before to go further we write down the right hand
side bracket expression of (4) in the following two equivalent ways:
\begin{equation}
\left[\left(\mathrm{rot}\,\mathbf{B}+\frac{\partial\mathbf{(-E)}}{\partial
\xi}\right)\times \mathbf{B}+
\mathbf{(-E)}\mathrm{div}\,\mathbf{(-E)}\right];\,                         
\left[\left(\mathrm{rot}\,\mathbf{(-B)}+\frac{\partial\mathbf{E}}{\partial
\xi}\right)\times\mathbf{(-B)}+\mathbf{E}\mathrm{div}\,\mathbf{E}\right].
\end{equation}
These last two expressions (5) can be considered as obtained from the left
hand side of (4) under the substitutions
$(\mathbf{E},\mathbf{B})\rightarrow(\mathbf{B},\mathbf{-E})$ and
$(\mathbf{E},\mathbf{B})\rightarrow(\mathbf{-B},\mathbf{E})$ respectively.
Hence, the field $(\mathbf{E},\mathbf{B})$ has always as a {\it partner}
one of the fields $(\mathbf{-B},\mathbf{E})$, or
$(\mathbf{B},\mathbf{-E})$.

This observation suggests the following view: {\it an adequate mathematical
representation of a time dependent free electromagnetic field requires a
collection of two fields} :
$\Big[(\mathbf{E},\mathbf{B});(\mathbf{-B},\mathbf{E})\Big]$, or
$\Big[(\mathbf{E},\mathbf{B});(\mathbf{B},\mathbf{-E})\Big]$. We could
also say that {\it a real free field consists of two interacting subsystems
(partners) described by two component-fields}, and each component-field
has electric and magnetic components, and
is
determined by the other through rotation-like transformation, in particular,
if $\mathbf{E}.\mathbf{B}=0$ then this rotation is to $\pm\pi/2$. This
view and relation (4) suggest, in turn, that {\it the intrinsic dynamics of
the real time-dependent electromagnetic fields could be considered as
accompanied by a local energy-momentum exchange between the corresponding
two component-fields}. Such a view suggests also that each of the two
component-fields of a real free field may keep locally its energy-momentum
if the inter-exchange is simultaneous and in equal quantities.

We are going now to interpret the equation (4) in accordance with the view
on equations of motion as stated in the Introduction. Our object of interest
$\Phi$, representing the integrity of a real electromagnetic field, is the
couple $\Big[(\mathbf{E},\mathbf{B});(\mathbf{-B},\mathbf{E})\Big]$ (the
other case $\Big[(\mathbf{E},\mathbf{B});(\mathbf{B},\mathbf{-E})\Big]$ is
considered analogically). In view of the above considerations our equations
should be able to express first, internal for each component-field
energy-momentum redistribution, second, possible and admissible
energy-momentum exchange between the two component-fields. Hence, we have to
define the corresponding change-objects $D(\mathbf{E},\mathbf{B})$ and
$D(\mathbf{-B},\mathbf{E})$ for each component-field and their "projections"
on the two component-fields.

The change object $D(\mathbf{E},\mathbf{B})$
for the first component-field $(\mathbf{E},\mathbf{B})$ we, naturally, define as
$$
D(\mathbf{E},\mathbf{B}):=
\left(\mathrm{rot}\mathbf{E}+ \frac{\partial\mathbf{B}}{\partial \xi}; \,
\mathrm{div}\mathbf{B}\right).
$$
The corresponding "projection" of $D(\mathbf{E},\mathbf{B})$ on
$(\mathbf{E},\mathbf{B})$
$$
\mathfrak{P}\left[D(\mathbf{E},\mathbf{B});
(\mathbf{E},\mathbf{B})\right]=
\mathfrak{P}\left[\left(\mathrm{rot}\mathbf{E}+
\frac{\partial\mathbf{B}}{\partial \xi};
\, \mathrm{div}\mathbf{B}\right); (\mathbf{E},\mathbf{B})\right]
$$
is suggested by the left hand side of (4) and we define it by :
$$
\mathfrak{P}\left[\left(\mathrm{rot}\mathbf{E}+
\frac{\partial\mathbf{B}}{\partial \xi};
\, \mathrm{div}\mathbf{B}\right); (\mathbf{E},\mathbf{B})\right]:=
\left(\mathrm{rot}\,\mathbf{E}+\frac{\partial\mathbf{B}}{\partial
\xi}\right)\times \mathbf{E}+\mathbf{B}\mathrm{div}\,\mathbf{B}.
$$
For the second component-field $(-\mathbf{B},\mathbf{E})$, following the same
procedure we obtain:
$$
\mathfrak{P}\left[D(\mathbf{-B},\mathbf{E});
(\mathbf{-B},\mathbf{E})\right]=
\mathfrak{P}\left[\left(\mathrm{rot}\mathbf{(-B)}+
\frac{\partial\mathbf{E}}{\partial \xi};
\, \mathrm{div}\mathbf{E}\right); (\mathbf{-B},\mathbf{E})\right]=
$$
$$
=\left(\mathrm{rot}\,\mathbf{(-B)}+\frac{\partial\mathbf{E}}{\partial
\xi}\right)\times \mathbf{(-B)}+\mathbf{E}\mathrm{div}\,\mathbf{E}=
\left(\mathrm{rot}\,\mathbf{B}-\frac{\partial\mathbf{E}}{\partial
\xi}\right)\times \mathbf{B}+\mathbf{E}\mathrm{div}\,\mathbf{E}.
$$
Hence, relation (4) looks like
$$
\mathfrak{P}\left[D(\mathbf{E},\mathbf{B});
(\mathbf{E},\mathbf{B})\right]+
\mathfrak{P}\left[D(\mathbf{-B},\mathbf{E});
(\mathbf{-B},\mathbf{E})\right]=0 .
$$

The accepted two-component view on a real time dependent electromagnetic field
allows in principle admissible energy-momentum exchange with the outside world
through any of the two component-fields. Hence, the above calculations suggest
to interpret the two sides of (4) as momentum quantities that each
component-field $(\mathbf{E},\mathbf{B})$, or $(\mathbf{-B},\mathbf{E})$, is
potentially able to give to some other physical object and these quantities are
expressed in terms of $\mathbf{E},\mathbf{B}$ and their derivatives only. In
the case of free field, since no momentum is lost by the field, there are two
possibilities: each component-field to keep the energy-momentum it carries, or
one of the component-fields to change its energy-momentum at the expense of the
other. If we denote by $\Delta_{11}$ and by $\Delta_{22}$ the allowed
energy-momentum changes of the two component fields, by $\Delta_{12}$ the
energy-momentum that the first component-field receives from the second
component-field, and by $\Delta_{21}$ the energy-momentum that the second
component-field receives from the first component-field, then according to the
energy-momentum local conservation law we may write the following equations:
\[
\Delta_{11}=\Delta_{12}+\Delta_{21}; \ \
\Delta_{22}=-\left(\Delta_{21}+\Delta_{12}\right),
\]
which is in accordance with the equation (4):
$\Delta_{11}+\Delta_{22}=0$.

We determine now how the mutual momentum exchange between the two
component-fields
$\mathbf{P}_{(\mathbf{E},\mathbf{B})}\rightleftarrows
\mathbf{P}_{(\mathbf{-B},\mathbf{E})}$,
or,
$\mathbf{P}_{(\mathbf{E},\mathbf{B})}\rightleftarrows
\mathbf{P}_{(\mathbf{B},\mathbf{-E})}$
is performed, i.e. the explicit expressions for $\Delta_{12}$ and
$\Delta_{21}$.
The formal expressions are easy to obtain. In fact, in the case
$\mathbf{P}_{(\mathbf{E},\mathbf{B})}\rightarrow
\mathbf{P}_{(\mathbf{-B},\mathbf{E})}$, i.e. the quantity $\Delta_{21}$,
we have to "project" the change object for the second component-field
$$
D(\mathbf{-B},\mathbf{E}):=
\left(\mathrm{rot}\mathbf{(-B)}+ \frac{\partial\mathbf{E}}{\partial \xi}; \,
\mathrm{div}\mathbf{E}\right)
$$
on the first component-field $(\mathbf{E},\mathbf{B})$. We obtain:
\begin{equation}
\Delta_{21}=
\left(\mathrm{rot}\,(\mathbf{-B})+\frac{\partial\mathbf{E}}{\partial
\xi}\right)\times \mathbf{E}+\mathbf{B}\mathrm{div}\,\mathbf{E}=
-\left(\mathrm{rot}\,\mathbf{B}-\frac{\partial\mathbf{E}}{\partial
\xi}\right)\times \mathbf{E}+\mathbf{B}\mathrm{div}\,\mathbf{E} \ .
\end{equation}
In the reverse case
$\mathbf{P}_{(\mathbf{-B},\mathbf{E})}\rightarrow
\mathbf{P}_{(\mathbf{E},\mathbf{B})}$, i.e. the case $\Delta_{12}$,
we have to project the change-object
for the first component-field given by
$$
D(\mathbf{E},\mathbf{B}):=
\left(\mathrm{rot}\mathbf{E}+ \frac{\partial\mathbf{B}}{\partial \xi}; \,
\mathrm{div}\mathbf{B}\right)
$$
on the second component-field $(\mathbf{-B},\mathbf{E})$.
We obtain
\begin{equation}
\Delta_{12}=
\left(\mathrm{rot}\,\mathbf{E}+\frac{\partial\mathbf{B}}{\partial
\xi}\right)\times (\mathbf{-B})+\mathbf{E}\mathrm{div}\,\mathbf{B}=
-\left(\mathrm{rot}\,\mathbf{E}+\frac{\partial\mathbf{B}}{\partial
\xi}\right)\times\mathbf{B}+\mathbf{E}\mathrm{div}\,\mathbf{B}.
\end{equation}
So, the internal local momentum balance is governed by the equations
\vskip 0.2cm
\begin{equation}
\left(\mathrm{rot}\,\mathbf{E}+\frac{\partial\mathbf{B}}{\partial
\xi}\right)\times \mathbf{E}+\mathbf{B}\mathrm{div}\,\mathbf{B}= \\
-\left(\mathrm{rot}\,\mathbf{E}+\frac{\partial\mathbf{B}}{\partial
\xi}\right)\times \mathbf{B}+\mathbf{E}\mathrm{div}\,\mathbf{B}-
\left(\mathrm{rot}\,\mathbf{B}-\frac{\partial\mathbf{E}}{\partial
\xi}\right)\times \mathbf{E}+\mathbf{B}\mathrm{div}\,\mathbf{E},
\end{equation}
\vskip 0.3cm
\begin{equation}
\left(\mathrm{rot}\,\mathbf{B}-\frac{\partial\mathbf{E}}{\partial
\xi}\right)\times \mathbf{B}+\mathbf{E}\mathrm{div}\,\mathbf{E}= \\
\left(\mathrm{rot}\,\mathbf{B}-\frac{\partial\mathbf{E}}{\partial
\xi}\right)\times \mathbf{E}-\mathbf{B}\mathrm{div}\,\mathbf{E}+
\left(\mathrm{rot}\,\mathbf{E}+\frac{\partial\mathbf{B}}{\partial
\xi}\right)\times \mathbf{B}-\mathbf{E}\mathrm{div}\,\mathbf{B}.
\end{equation}
\vskip 0.2cm
These two vector equations (8)-(9) we consider as natural Newton type
field equations. According to them the intrinsic dynamics of a free
electromagnetic field is described by two couples of vector fields,
$[(\mathbf{E},\mathbf{B}); (\mathbf{-B},\mathbf{E})]$, or
$[(\mathbf{E},\mathbf{B}); (\mathbf{B},\mathbf{-E})]$, and this intrinsic
dynamics could be interpreted as a direct momentum exchange
between two well defined subsystems mathematically described by these two
component-fields.

A further natural specilization of the above two vector equations (8)-(9)
could be made if we assume that this internal momentum exchange
realizes a dynamical equilibrium between the two component fields, i.e.
each component-field conserves its momentum : $\Delta_{11}=\Delta_{22}=0$.
In such a situation each
component-field loses as much as it gains, so,
equations (8)-(9) reduce to
\begin{equation}
\Delta_{11}\equiv\left(\mathrm{rot}\,\mathbf{E}+\frac{\partial\mathbf{B}}{\partial
\xi}\right)\times \mathbf{E}+\mathbf{B}\mathrm{div}\,\mathbf{B}=0,     
\end{equation}
\begin{equation}
\Delta_{22}\equiv
\left(\mathrm{rot}\,\mathbf{B}-\frac{\partial\mathbf{E}}{\partial      
\xi}\right)\times \mathbf{B}+\mathbf{E}\mathrm{div}\,\mathbf{E}=0,
\end{equation}
\begin{equation}
\left(\mathrm{rot}\,\mathbf{E}+\frac{\partial\mathbf{B}}{\partial
\xi}\right)\times \mathbf{B}-\mathbf{E}\mathrm{div}\,\mathbf{B}+        
\left(\mathrm{rot}\,\mathbf{B}-\frac{\partial\mathbf{E}}{\partial
\xi}\right)\times \mathbf{E}-\mathbf{B}\mathrm{div}\,\mathbf{E}=0.
\end{equation}
Equation (12) fixes, namely, that {\it the exchange of
momentum density between the two component-fields is {\bf simultanious}
and in {\bf equal} quantities}, i.e. a {\bf permanent dynamical equilibrium}
between the two component-fields holds:
$\mathbf{P}_{(\mathbf{E},\mathbf{B})}\rightleftarrows
\mathbf{P}_{(\mathbf{-B},\mathbf{E})}$, or,
$\mathbf{P}_{(\mathbf{E},\mathbf{B})}\rightleftarrows
\mathbf{P}_{(\mathbf{B},\mathbf{-E})}$.

Note that, if equations (10) and (11) may be considered as
field-equivalents to the zero force field (eqn. (11)) and its dual (eqn.
(10)), this double-field viewpoint and the corresponding mutual
energy-momentum exchange described by equation (12) are essentially new
moments.

Equations (10)-(12) also suggest that the corresponding fields are able to
exchange energy-momentum with other physical systems in three ways. If
such an exchange has been accomplished, then the exchanged energy-momentum
quantities can be given in terms of the characteristics of the other
physical system (or in terms of the characteristics of the both systems,
e.g. the Lorentz force
$\rho\mathbf{E}+\frac{\mathbf{j}}{c}\times\mathbf{B}$), and to be
correspondingly equalized to the left hand sides of equations (10)-(12) in
accordance with the local energy-momentum conservation law.

Finally, we give the 4-dimensional relativistic picture (details see in [7]).
If the Minkowski pseudometric $\eta$ has
signature $(-,-,-,+)$ and
$F_{i4}=\mathbf{E}^i$, $F_{12}=\mathbf{B}^3,\,F_{13}=
-\mathbf{B}^2,\,F_{23}=\mathbf{B}^1$, and
$(*F)_{\alpha\beta}=-\frac12\varepsilon_{\alpha\beta\mu\nu}F^{\mu\nu}$,
$\mathbf{d}$ is the exterior derivative, $\delta=*\,\mathbf{d}*$ is the
coderivative, then the Maxwell stress tensor and its divergence are extended
respectively to $$ T_\mu^\nu=-\frac12\Big[F_{\mu\sigma}F^{\nu\sigma}+
(*F)_{\mu\sigma}(*F)^{\nu\sigma}\Big] \ \  \text{and} \ \
\nabla_\nu T_\mu^\nu=F_{\mu\nu}(\delta F)^\nu+(*F)_{\mu\nu}(\delta *F)^\nu.
$$
These expressions clearly and respectfully show the two-component
$(F,*F)$-structure of the field. Equations (8)-(9) are extended
correspondingly to
$$
(*F)_{\mu\nu}(\delta *F)^\nu=- \Big[F_{\mu\nu}(\delta *F)^\nu +
(*F)_{\mu\nu}(\delta F)^\nu\Big],
$$
$$ F_{\mu\nu}(\delta F)^\nu=
   F_{\mu\nu}(\delta *F)^\nu + (*F)_{\mu\nu}(\delta F)^\nu.
$$
Equations (10)-(12)
are extended correspondingly to
$$
 F^{\alpha\beta}(\mathbf{d}F)_{\alpha\beta\mu}\equiv
(*F)_{\mu\nu}(\delta *F)^\nu=0,  \ \ \
(*F)^{\alpha\beta}(\mathbf{d}*F)_{\alpha\beta\mu}\equiv
F_{\mu\nu}(\delta F)^\nu=0, \ \ \alpha<\beta;
$$
$$
(*F)^{\alpha\beta}(\mathbf{d}F)_{\alpha\beta\mu}+
F^{\alpha\beta}(\mathbf{d}*F)_{\alpha\beta\mu}\equiv
(\delta *F)^\nu F_{\nu\mu}+(\delta F)^\nu (*F)_{\nu\mu}=0, \ \ \alpha<\beta.
$$
\section{Some Properties of the nonlinear solutions}
Clearly, all solutions to
Maxwell pure field equations (*) are solutions to our nonlinear equations
(8)-(9) and (10)-(12), we shall call these solutions linear, and will not be
interested of them. In this section we shall concentrate on those solutions of
(10)-(12) which satisfy the conditions
\[
\mathrm{rot}\,\mathbf{E}+\frac{\partial\mathbf{B}}{\partial \xi}\neq 0,\quad
\mathrm{rot}\,\mathbf{B}-\frac{\partial\mathbf{E}}{\partial \xi}\neq 0,\quad
\mathrm{div}\,\mathbf{E}\neq 0,\quad \mathrm{div}\,\mathbf{B}\neq 0.
\]
These solutions we call further nonlinear. We note some of the
properties they have.
\vskip 0.3cm
$1.\ \mathbf{E}.\mathbf{B}=0;$
\vskip 0.3cm
$2. \
\ \left(\mathrm{rot}\,\mathbf{E}+
\frac{\partial\mathbf{B}}{\partial \xi}\right).\mathbf{B}=0$; \ \
$\left(\mathrm{rot}\,\mathbf{B}-
\frac{\partial\mathbf{E}}{\partial \xi}\right).\mathbf{E}=0$, \ \
From these two relations the classical Poynting energy-momentum
balance equation follows.
 \vskip 0.3cm
The above two properties are obvious from equations (10) and (11).
 \vskip 0.3cm
3. If $(\mathbf{E},\mathbf{B})$ defines a solution then
$(\mathbf{E}',\mathbf{B}')=
(a\mathbf{E}-b\mathbf{B};\  b\mathbf{E}+a\mathbf{B})$,
where $a,b\in\mathbb{R}$, defines also a solution. This property is
immediately verified through substitution.
 \vskip 0.3cm
4. $\mathbf{E}^2=\mathbf{B}^2$. To prove this, we first multiply equation (10)
on the left by $\mathbf{E}$ and equation (11) by
$\mathbf{B}$ (scalar products). Then we make use of the above properties 1
and 2 and of the vector algebra relation $X.(Y\times Z)=Z.(X\times Y)$.

Properties (1) and (4) say that all nonlinear solutions to (10)-(12) are {\it
null fields}, i.e. the two well known relativistic invariants
$I_1=\mathbf{B}^2-\mathbf{E}^2$ and $I_2=2\mathbf{E}.\mathbf{B}$ of the
 field are zero. \vskip 0.3cm 5.\ \
$\mathbf{B}.\left(\mathrm{rot}\,\mathbf{B}- \frac{\partial\mathbf{E}}{\partial
\xi}\right)- \mathbf{E}. \left(\mathrm{rot}\,\mathbf{E}+
\frac{\partial\mathbf{B}}{\partial \xi}\right)=
\mathbf{B}.\mathrm{rot}\mathbf{B}-\mathbf{E}.\mathrm{rot}\mathbf{E}=0.$
\vskip 0.3cm
\noindent
To prove this property we first multiply (vector product) (10) from the
right by $\mathbf{E}$, recall property 1, then multiply (scalar product)
from the left by $\mathbf{E}$, recall again $\mathbf{E}.\mathbf{B}=0$,
then multiply from the right (scalar product) by $\mathbf{B}$ and recall
property 4.

Property (5) suggests the following consideration. If $\mathbf{V}$ is an
arbitrary vector field on $\mathbb{R}^3$ then the quantity
$\mathbf{V}.\mathrm{rot}\mathbf{V}$ is known as {\it local helicity} and
its integral over the whole region occupied by $\mathbf{V}$ is known as
{\it integral helicity}, or just as {\it helicity} of $\mathbf{V}$.
Hence, property 5 says that the electric and magnetic components of a
nonlinear solution generate the same helicities. If we consider (through
the euclidean metric) $\mathbf{E}$ as 1-form on $\mathbb{R}^3$ and denote
by $\mathbf{d}$ the exterior derivative, then
$\mathbf{E}\wedge\mathbf{d}\mathbf{E}=
\mathbf{E}.\mathrm{rot}\mathbf{E}\,dx\wedge dy\wedge dz$, so, the zero
helicity says that the 1-form $\mathbf{E}$ defines a completely integrable
Pfaff system. The nonzero helicity says that the 1-form $\mathbf{E}$
defines non-integrable 1d Pfaff system, so the nonzero helicity defines
corresponding curvature. Therefore the equality between the
$\mathbf{E}$-helicity and the $\mathbf{B}$-helicity suggests to consider
the corresponding integral helicities
$\int_{\mathbb{R}^3}\mathbf{E}\wedge\mathbf{d}\mathbf{E}
=\int_{\mathbb{R}^3}\mathbf{B}\wedge\mathbf{d}\mathbf{B}$ (when they take
finite nonzero values) as a measure of the spin properties of the
solution.
\vskip 0.3cm
6.\ \ Example of nonlinear solution:
\begin{align*}
&\mathbf{E}=\left[\phi(x,y,ct+\varepsilon z)
\mathrm{cos}(-\kappa\frac{z}{l_o}+const), \,
\phi(x,y,ct+\varepsilon z)\mathrm{sin}(-\kappa\frac{z}{l_o}+const),\,0\right];\\
&\mathbf{B}=\left[\varepsilon \phi(x,y,ct+\varepsilon z)\,
\mathrm{sin}(-\kappa\frac{z}{l_o}+const),\,
-\varepsilon \phi(x,y,ct+\varepsilon z)
\mathrm{cos}(-\kappa\frac{z}{l_o}+const),\,0\right],
\end{align*}
where $\phi(x,y,ct+\varepsilon z)$ is an arbitrary positive function,
$l_o$ is an arbitrary positive constant with physical dimension of length,
and $\varepsilon$ and $\kappa$ take values $\pm1$ independently. The form
of this solution shows that the initial condition is determined entirely
by the choice of $\phi$, and it suggests also to choose the initial
condition $\phi_{t=0}(x,y,\varepsilon z)$ in the following way. Let for
$z=0$ the initial condition $\phi_{t=0}(x,y,0)$ be located on a disk
$D=D(x,y;a,b;r_o)$ of small radius $r_o$, the center of the disk to have
coordinates $(a,b)$, and the value of $\phi_{t=0}(x,y,0)$ to be
proportional to the distance $R(x,y,0)$ between the origin of the
coordinate system and the point $(x,y,0)$, so, $R(x,y,0)=\sqrt{x^2+y^2}$,
and $D$ is defined by $D=\{(x,y)|\sqrt{(x-a)^2+(y-b)^2}\leq r_o\}$. Also,
let $\theta_D$ be the smoothed out characteristic function of the disk
$D$, i.e. $\theta_D=1$ everywhere on $D$ except a very thin hoop-like zone
$B_D\subset D$ close to the boundary of $D$ where $\theta_D$ rapidly goes
from 1 to zero (in a smooth way), and $\theta_D=0$ outside $D$. Let also
the dependence on $z$ to be given by
be the corresponding characteristic function $\theta(z;2\pi l_o)$ of an
interval $(z,z+2\pi l_o)$ of length $2\pi l_o$ on the $z$-axis. If
$\gamma$ is the proportionality coefficient we obtain
$$
\phi(x,y,z,ct+\varepsilon z)=
\gamma.R(x,y,0).\theta_D.\theta(ct+\varepsilon z;2\pi l_o).
$$
We see that because of the available {\it sine} and {\it cosine} factors in the
solution, the initial condition for the solution will occupy a helical cylinder
of height $2\pi l_o$, having internal radius of $r_o$ and wraped up around the
$z$-axis. Also, its center will always be $R(a,b,0)$-distant from the $z$-axis.
Hence, the solution will propagate translationally - along the coordinate $z$
with the velocity $c$, and rotationally - inside the corresponding infinitely
long helical cylinder because of the $z$-dependence of the available periodical
multiples. The curvature $K$ and the torsion $T$ of the screwline through the
point $(x,y,0)\in D$ will be
\[
K=\frac{R(x,y,0)}{R^2(x,y,0)+l_o^2},\ \ \ \
T=\frac{\kappa l_o}{R^2(x,y,0)+l_o^2} \ .
\]
The rotational frequency $\nu$ will be $\nu=c/2\pi l_o$, so we can introduce
the period $T=1/\nu$ and elementary action $h=E.T$, where $E$ is the (obviously
finite) integral energy of the solution defined as 3d-integral of the energy
density $(\mathbf{E}^2+\mathbf{B}^2)/2=\phi^2$ (see the figures on p.62 in {\it
hep-th}/0403244).

This example presents also a completely integrable differential (2-dimensional)
system, i.e. there exist two functions $f$ and $g$ such that the Lie bracket
$[\mathbf{E},\mathbf{B}]$ can be represented in the form
$f\,\mathbf{E}+g\,\mathbf{B}$. The appropriately normed local helicities
$\frac{2\pi l_o^2}{c}\mathbf{E}\wedge\mathbf{d}\mathbf{E}= \frac{2\pi
l_o^2}{c}\mathbf{B}\wedge\mathbf{d}\mathbf{B}$ generate the integral helicity
$h=E.T$, i.e. the elementary action, where $E$ is the integral energy of the
solution, $T=2\pi l_o/c$, and, clearly, $h=const$. If we interpret $h$ as the
Planck's constant then the relation $h=E.T$ is equivalent to the Planck's
relation $E=h\nu$, and $h$ appears as the (integral) spin of the solution.

\section{Discussion and Conclusion}
The main idea of the paper is that carrying out the Newton way for writing down
dynamical equations for particles in mechanics to writing down dynamical
equations for continuous field systems should naturally result to nonlinear
partial differential equations even in non-relativistic theories. Moreover,
clarifying the sense of the information included in these dynamical equations
according to the Newton approach, we come to the conclusion formulated in the
Introduction, namely, we have to mathematically describe those changes of the
object considered which are qualified as {\it admissible} and {\it consistent}
with the system's identification and with the local energy-momentum balance
relations.  In the case of "free" systems these relations represent the local
energy-momentum conservation properties of the system.  The energy-momentum
characteristics are chosen because of their two important properties: they are
physically {\it universal} and {\it conservative}. This means that {\it every}
physical object carries nonzero energy-momentum and, vice versa, {\it every}
quantity of energy-momentum is carried by some physical object. Also, if a
physical object loses/gains some quantity of energy-momentum then some other
physical object necessarily gains/loses the same quantity of energy-momentum.
If this viewpoint is assumed, then the problem of finding appropriate dynamical
equations for an object reduces mainly to: {\bf first}, getting knowledge of
the potential abilities of the object considered to lose and gain
energy-momentum; {\bf second}, to create adequate mathematical quantities
describing locally these abilities.

The electromagnetic field, considered as a continuous physical object of
special kind, gives a good example in this direction since, thanks to
Maxwell's fundamental and summarizing works, all the information
needed is available. The notices of Poynting [5], and Thomson, Poincare and
Abraham [6], showing the importance of the (deduced from Maxwell equations)
vector $\frac1c\mathbf{E}\times\mathbf{B}$ from local energy-momentum
propagation point of view, has completed the resource of adequate
and appropriate mathematical objects since it appears as natural complement of
Maxwell stress tensor, and allows to write down dynamical field equations
having direct local energy-momentum balance sense. However, looking back in
time, we see that this viewpoint for writing down field equations has been
neglected, theorists have paid more respect and attention to the "linear part"
of Maxwell theory, enjoying, for example, the {\it exact} but not realistic,
and even {\it physically senseless} in many respects, plane wave solutions in
the pure field case.

Therefore, not so long after the appearance of Maxwell
equations the photoeffect experiments showed the nonadequateness of the linear
part of Maxwell theory as a mathematical model of electromagnetic fields
producing realistic model-solutions of free time-dependent fields. Although the
almost a century long time development of standard quantum and relativistic
quantum theories that followed, a reasonable model-solutions describing
individual photons, considered as basic, spatially finite and time-stable
objects, these theories have not presented so far. Nobody doubts nowadays that
photons really exist, and this very fact suggests to try first classical field
approach in finding equations admitting 3d-finite and time stable solutions
with appropriate properties.

The historical perspective suggests to follow the 4-potential approach,
but modern knowledge and experience, and even the Maxwell stress tensor
achievements, suggest some different views. In fact, we have all reasons
to consider the microobjects as real as all other physical objects, so, no
point-like charges and infinite field model-solutions should be considered
as adequate. Since the 4-potential approach in Maxwell theory
does not allow spatially
finite and time stable pure field solutions with photon-like structure and
behavior its interpretation as a basic concept does not seem to be
appreciable. Also, the 4-potential approach excludes many solutions of the
charge free Maxwell equations. For example, in relativistic terms the
Coulomb field is given by the 2-form $F=\frac{q}{r^2}dr\wedge d\xi, \
\mathbf{d}F=0$, its Minkowski-dual is $*F=q\sin\,\theta\, d\theta\wedge
d\varphi, \ \mathbf{d}*F=0$, where $F$ has a global 4-potential, but $*F$
has NO global 4-potential. Now, the 2-parameter family of 2-forms
$(\mathfrak{F},*\mathfrak{F})=(aF-b*F; bF+a*F), a,b\in\mathbb{R}$, gives
an infinite number of solutions to Maxwell equations
$\mathbf{d}\mathfrak{F}=0, \mathbf{d}*\mathfrak{F}=0$ admitting NO global
4-potential. This suggests the view that the 4-potential can be used as a
{\it working tool} (wherever it causes no controversies) but {\it not as a
basic concept}.

In conclusion, paying due respect to the Newton view on dynamical
equations and to the local energy-momentum conservation law we based our
approach on the Maxwell stress tensor and on the Poynting vector as
natural quantities carrying the physically meaningful energy-momentum
characteristics of the electromagnetic field. The natural description in these
terms is based on two component-fields:
$[(\mathbf{E},\mathbf{B})],[(\mathbf{-B},\mathbf{E})]$, or
$[(\mathbf{E},\mathbf{B})],[(\mathbf{B},\mathbf{-E})]$, and mutual
energy-momentum exchange between the two component-fields (in equal
quantities) is considered as possible and explicitly accounted. The
equations obtained include all solutions to the charge free Maxwell
equations. A basic part of the new (nonlinear) solutions have zero
invariants. Among these zero-invariant new solutions there are
time-stable and spatially finite ones having photon-like properties and
behavior. An analog of the Planck relation $E=h\nu$ holds for these
solutions, where the constant $h$ appears as an integral helicity of such
a solution.
\vskip 0.3cm
This study was partially supported by
Contract $\phi\,15\,15$ with the Bulgarian National Fund "Science Research".
\vskip 0.5cm
\newpage
{\bf REFERENCES}
\vskip 0.3cm

[1] {\bf Poisson, S. D}. {\it Mem. Acad. sci.}, vol.3, p.121 (1818)

[2] {\bf Courant, R., Hilbert, D}., {\it Methoden der mathematischen
Physik}, Berlin, vol.2 \S 6 (1937)

[3] {\bf Farlow, S. J}., {\it Partial Differential equations for
Scientists and Engineers}, John Wiley and Sons, Inc., 1982

[4] {\bf Maxwell, J. C}., On Physical Lines of Force. Part 1., {\it Phil.
Mag}. vol.XXI (1861), vol. XXIII (1862); also, {\it The Scientific Papers
of James Clerk Maxwell}, vol.I, pp.451-513 (1890)

[5] {\bf Poynting, J. H}., Phil. Trans. {\bf 175}, 1884, pp.343-361.

[6] {\bf Thomson, J.J.}, Recent Researches in Elect. and Mag., 1893, p.13;
{\bf Poincare, H}., Archives Neerland Sci., vol.{\bf 2}, 1900,
pp.252-278; {\bf Abraham, M.}, Gott.Nach., 1902, p.20;
see also the corresponding comments in Whitakker's {\it History of
the theories of Aether and Electricity}, vol.1, Ch.10.

[7] {\bf Donev, S., Tashkova, M}., {\it Proc. Roy. Soc. of London} A 450,
281 (1995), see also: \newline hep-th/0403244 .

\end{document}